 \documentclass[final,number,3p,times,twocolumn]{elsarticle}

\usepackage{amssymb}

\usepackage{mathptmx}      
\usepackage{latexsym}
\usepackage{graphicx}
\usepackage{graphics}
\usepackage{epsfig}
\usepackage[latin1]{inputenc}
\usepackage{latexsym}
\usepackage{amsmath}
\usepackage{enumerate}
\usepackage{natbib}

\def\ket#1{\lvert#1\rangle}

\journal{Physica B}

\begin{document}

\begin{frontmatter}



\title{Energy entanglement in neutron interferometry}


\author{Katharina Durstberger-Rennhofer}
\ead{Durstberger@ati.ac.at}
\author{Yuji Hasegawa}
\ead{Hasegawa@ati.ac.at}

\address{Atominstitut, Technische Universit\"{a}t Wien, Stadionallee 2, A-1020 Wien, Austria  }

\begin{abstract}

Entanglement between degrees of freedom, namely between the spin, path and (total) energy
degrees of freedom, for single neutrons is exploited. We implemented a triply entangled
Greenberger-Horne-Zeilinger(GHZ)-like state and coherently manipulated relative phases of
two-level quantum subsystems. An inequality derived by Mermin was applied to analyze the generated  GHZ-like state: we determined the four expectation values and finally obtained $M=2.558\pm0.004\nleq2$. This demonstrates the violation of a Mermin-like inequality for triply entangled GHZ-like state in a single-particle system, which, in turn, exhibits a clear inconsistency between noncontextual assumptions and quantum mechanics and confirms quantum contextuality.

\end{abstract}


\end{frontmatter}


\section{Introduction}

It was Einstein, Podolsky, and Rosen (EPR) \cite{EPR} and afterwards Bell \cite{Bell} who shed light on the non-local properties between subsystems in quantum mechanics.
Bell inequalities \cite{Bell} are constraints imposed by local hidden-variable theories (LHVTs) on the values of some specific linear combinations of the averages of the results of spacelike separated experiments on distant systems. Reported experimental violations of Bell inequalities, e.g., with photons \cite{WJSWZ98} or atoms \cite{Matsukevich}, suggest that quantum mechanics (QM) cannot be reproduced by LHVTs.

In single particle systems where different degrees of freedom (DOFs) are entangled, Bell-like inequalities can be tested. In this scenario, the conflict arises not between QM and LHVTs but a violation confirms the impossibility of noncontextual hidden variable theories (NCHVTs) \cite{Mermin93}. Experimental violations of Bell-like inequalities can be found for single neutrons \cite{Hasegawa03} (entanglement between spin and path DOF) and for single photons \cite{Amselem09} (entanglement between polarization and path DOF) confirming quantum contextuality.

Kochen and Specker \cite{KS67} were the first to analyse the concept of contextuality in QM which is a more general concept than non-locality, and leads to striking phenomena predicted by quantum theory.

The conflict between LHVTs and QM is even more apparent in tri- or multipartite quantum systems which was analysed by Greenberger, Horne and Zeilinger (GHZ) \cite{GHZ,GHSZ}. Here the contradiction leads to nonstatistical predictions in contrast to common Bell-inequalities.
Mermin \cite{Mermin90b} showed that this conflict can be converted into a larger violation of a Bell-like inequality between three or more separated systems. Experimental tests of these inequalities were reported, e.g., for three and four photons \cite{Pan,Zhao} as well as four ions \cite{Sackett}.

Apart from technical challenges of preparing a GHZ-like entangled state for single neutrons, the violation of a Mermin-like inequality is interesting in itself because it is more robust to noise or disturbances than previous reported violations of Bell-like inequalities \cite{Hasegawa03} which emphasizes the conflict between QM and NCHVTs.

Here, we describe the experimental realization of tripartite entanglement for single neutrons \cite{Hasegawa2010} where one external degree of freedom (path states in the interferometer) is entangled with two internal degrees of freedom (spin and energy) leading to a violation of a Mermin-like inequality \cite{Mermin90b}.

\section{Tripartite entanglement: GHZ argument}

The GHZ state was first proposed for four spin-$1/2$ particles by Greenberger, Horne and Zeilinger \cite{GHZ}. Later on Mermin \cite{Mermin90a} presented a version with three spin-$1/2$ particles; in Ref. \cite{Mermin90} he pointed out how GHZ-states can be used to reveal the relation between Kochen-Specker (KS) theorem \cite{KS67} and Bells theorem \cite{Bell}.

Suppose three spin-$1/2$ particles emitted from a common source in the so called GHZ-state,\begin{equation}
	\lvert\psi^{GHZ}\rangle=\frac{1}{\sqrt2}(\lvert\uparrow\uparrow\uparrow\rangle
		-\lvert\downarrow\downarrow\downarrow\rangle)\,,
\end{equation}
fly apart in different directions. Measurements on the three particles are described by Pauli-spin operators with measurement outcomes $\pm 1$.
Consider the following three hermitian, commuting operators
\begin{equation} \label{eq:meas.op}
	\{A_i\}=\{\sigma_x^a \sigma_y^b \sigma_y^c \;,\;
	\sigma_y^a \sigma_x^b \sigma_y^c \;,\;
	\sigma_y^a \sigma_y^b \sigma_x^c \}\,,
\end{equation}
which satisfy the eigenvalue equation
\begin{equation}\label{eq:eigenstate,xyy}
	A_i \lvert\psi^{GHZ}\rangle=+1\lvert\psi^{GHZ}\rangle\,.
\end{equation}
For a system prepared in state $\lvert\psi^{GHZ}\rangle$ we can measure, for example, the x-component of particle $a$ and the y-component of particle $b$ and infer the result of the y-measurement of particle $c$ by the fact that the product of all three measurements is $+1$. The same reasoning holds for the other operators in Eq. (\ref{eq:meas.op}). Thus we may conclude that the system is characterized by predefined number, $m_x^a$, $m_y^a$, $m_x^b$, $m_y^b$, $m_x^c$, and $m_y^c$ with values $\pm1$, representing the measurement outcomes.

The conflict arises if we measure the operator
$\sigma_x^a \sigma_x^b \sigma_x^c=-A_1\cdot A_2\cdot A_3$
which commutes with all $A_i$ and satisfies the eigenvalue equation
\begin{equation}\label{eq:eigenstate,xxx}
	\sigma_x^a \sigma_x^b \sigma_x^c\lvert\psi^{GHZ}\rangle=-1\lvert\psi^{GHZ}\rangle\,.
\end{equation}
The measurement outcomes have to obey the relations
\begin{equation}\label{eq:relations}
\begin{split}
	m_x^a m_y^b m_y^c&=1 \,,\qquad m_y^a m_x^b m_y^c=1\,,\\
	m_y^a m_y^b m_x^c&=1 \,, \qquad m_x^a m_x^b m_x^c=-1\,, \\
\end{split}
\end{equation}
which turns out to be impossible. By multiplication of the left hand sides of (\ref{eq:relations}) as well as the right hand sides we get $+1$ for the left hand sides because all outcomes appear twice whereas the right hand side gives $-1$.
The argument leads to a contradiction because the perfect correlations of QM (perfect predictability of measurement outcomes) are incompatible with the premisses of LHVT (locality and reality). The above reasoning also can be cast into a KS form which is independent of specific states (details may be found in Ref. \cite{Mermin90}).

\section{GHZ-like entanglement for single neutrons}

\subsection{Coherent energy manipulation}

\begin{figure}[h!]
\begin{center}
\includegraphics[width=0.4\textwidth]{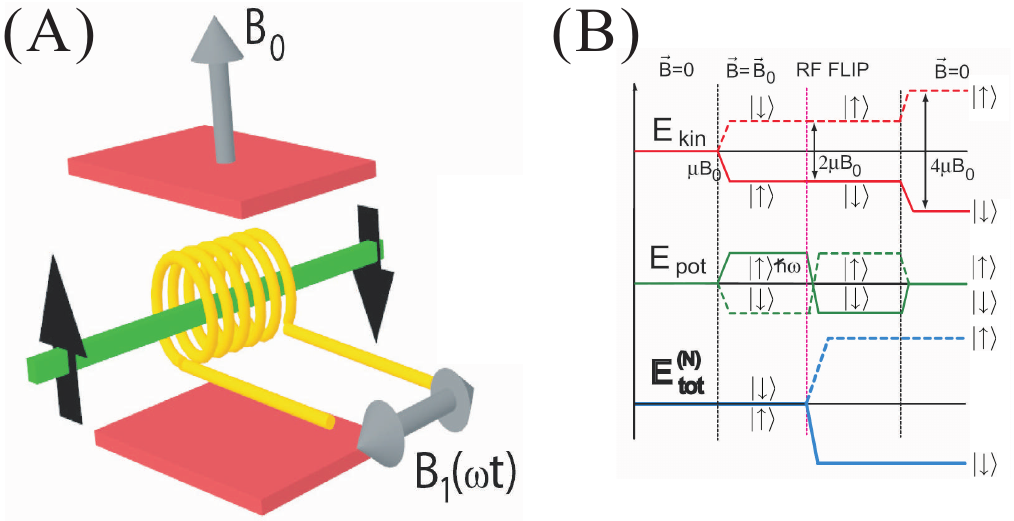}
\caption{ Coherent energy manipulation scheme by the interaction with an oscillating magnetic field. (A) Spin-flip process through the time-dependent magnetic field, $B_1(\omega t)$ (B) Energy diagram, i.e., kinetic, potential, and total energies, of neutrons in passing through the magnetic field configuration of a RF-flipper.}
\label{Fig:1}
\end{center}
\end{figure}

In a radio-frequency (RF) oscillating magnetic field the incoming neutron changes the spin direction as well as the total energy. The scheme of coherent energy-manipulation \cite{Sponar08a} is depicted in Fig. \ref{Fig:1}.

A polarized neutron enters an area of a spatially-distributed guide magnetic field, $B_0$, which induces a shift of potential-energy, $\Delta E_{pot}=\pm \mu B_0$ ($\pm$ correspond to parallel and anti-parallel spin-states and $\mu$ is the neutron magnetic moment) due to the Zeeman effect. This is accompanied by a change of kinetic energy accordingly: the total energy, given by the sum of kinetic and potential energies, is conserved. Then, the spin is flipped by an additional oscillating magnetic field $B_{1}(\omega t)= B_1 cos(\omega t+\phi)$ in a constant guide field $B_0$, which, in turn, leads to a coherent (total) energy manipulation. In contrast to the time-independent spatial interaction of $\mu B_0$, a time-dependent interaction induces changes in the potential energy and the kinetic energy is kept constant. It is worth noting here that, although the spin flip occurs in manipulating the total energy, the spin itself can be flipped, or rather arbitrarily manipulated, independently with the use of stationary magnetic field, e.g., by a DC-coil. This fact implies that a spin-flip by a RF-flipper accompanied by another spin-flip by a DC-flipper afterwards or beforehand effectively works as a (total) energy manipulation without altering the spin: the spin, and energy degrees of freedoms (DOFs) in our experiments are independently manipulable.

\subsection{Tripartite entanglement: GHZ-like state}

In a perfect crystal neutron interferometer experiment \cite{Rauch00}, the up-polarized incident neutron beam, denoted by $\ket{\uparrow}$, passes
through the beam-splitter plate of the interferometer, thereby the state describing neutron's path is transformed into a superposition of path states, $\frac{1}{\sqrt{2}} (\ket{\textrm{I}}+\ket{\textrm{II}})$.
In the interferometer, a RF spin-flipper is inserted in the path II, where the spin-flip process by a time-dependent interaction induces energy transitions from the initial energy state $\ket{E_0}$ to states $\ket{E_0\!-\!\hbar\omega}$ by photon exchange \cite{Summi93}. Consequently, one can generate neutrons in a triply entangled GHZ-like state, given by
\begin{equation}
 \label{eq:eq1}
\ket{\Psi^{GHZ}_N}=\frac{1}{\sqrt{2}}\Bigl(\ket{\uparrow} \otimes \ket{\textrm{I}} \otimes \ket{E_0} + \ket{\downarrow} \otimes \ket{\textrm{II}} \otimes \ket{E_0-\hbar\omega} \Bigr).
\end{equation}

\begin{figure}
\begin{center}
   \includegraphics[width=0.48\textwidth]{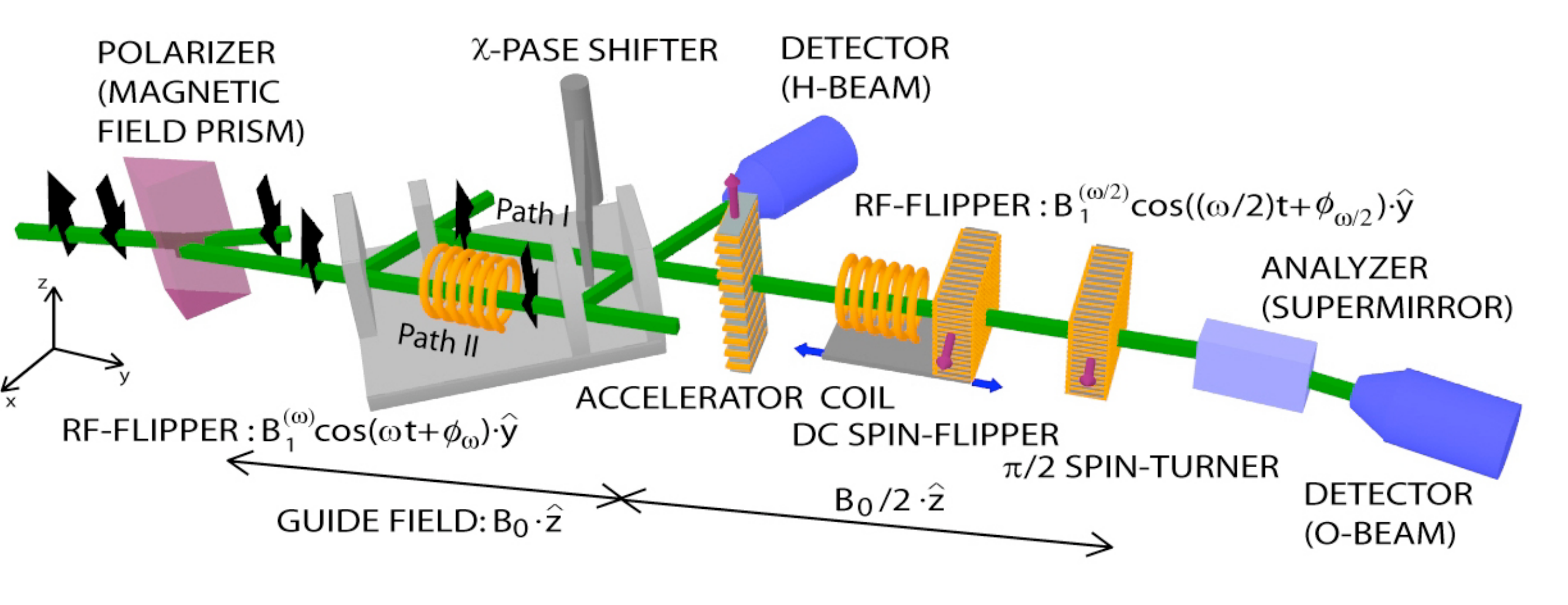}
   \caption{Experimental setup for the study of tripartite entangled GHZ-like state in a single-neutron system.}
\label{Setup} \end{center}
\end{figure}

The state of neutrons is characterized by three (the spin, path and energy) DOFs: all of them are described simply by two-level quantum systems. Measurement operators $\sigma_x$ and $\sigma_y$ in each DOF are accomplished by phase manipulations between superpositions of the basis states in each DOF.

(i) The spin-phase $\alpha$ is adjusted by a magnetic field oriented along +z direction ('accelerator' coil) together with a DC-flipper in $\pi/2$-flipping mode.

(ii) The phase manipulation of the path DOF is accomplished with an auxiliary phase shifter $\chi$ made of a parallel-sided Si plate 5mm in thickness and the last plate of the interferometer.

(iii) The so-called zero-field precession phase $\gamma$ \cite{Golub94} is employed for the phase manipulation of the energy DOF; the second RF-flipper together with a DC-flipper are used. (An experimentally convenient method to manipulate individually the Larmor phase $\alpha$ and the zero-field phase $\gamma$ was found and reported in \cite{Sponar08b}.)

\subsection{Mermin-like inequality}

Mermin \cite{Mermin90b} analyzed the GHZ argument in detail and derived an inequality suitable for experimental tests to distinguish between predictions of QM and LHVTs  since perfect correlations cannot be observed in real experiments (see also Ref. \cite{Cabello2002}).
In a similar way, assuming a tripartite system and taking the assumption in the conditionally independent form due to NCHVTs instead of LHVTs, the border for a sum of expectation values of certain product observables is obtained. The sum of expectation values of product observables, called $M$, is defined as
\begin{equation}
 \label{eq:eq3}
\begin{split}
M\!=\!E[\sigma^s_x \sigma^p_x  \sigma^e_x]
-\!E[\sigma^s_x  \sigma^p_y  \sigma^e_y]\\
-\!E[\sigma^s_y  \sigma^p_x  \sigma^e_y]
-\!E[\sigma^s_y  \sigma^p_y  \sigma^e_x]
\end{split}
\end{equation}
where $\!E[\ldots]$ represents expectation values, and  $\sigma^s_j$, $\sigma^p_j$, and $\sigma^e_j$ represent Pauli operators for the two-level systems in the spin, path, and energy DOF, respectively. NCHVTs demands $|M| \leq 2$, while quantum theory predicts an upper bound of 4: any measured value of $M$ that is larger than 2 decides in favor of quantum contextuality.

\section{Neutron interferometric experiments}

	The experiment was carried out at the neutron-interferometer beam line S18 at the high flux reactor at the Institute Laue Langevin (ILL). A schematic view of the experimental setup is shown in Fig. \ref{Setup}. Magnetic prisms were used to polarize the incident beam vertically, and the interferometer was adjusted to give the 220 reflections. A parallel-sided Si plate was used as a phase shifter to tune the phase $\chi$ for the path DOF. The first RF spin-flipper was located in a fairly uniform magnetic guide field, and its operational frequency was tuned to $\omega=\text{58kHz}$. The GHZ-like state of neutrons $\ket{\Psi^{GHZ}_N}$ was generated by turning on this RF spin-flipper. The second RF spin-flipper, tuned to the operational frequency $\omega/2=\text{29kHz}$, was placed in the region downstream of the interfering O-beam. This RF spin-flipper was mounted on a common translator together with a DC spin-flipper. The translation of the common basis allows to tune the phase $\gamma$ of the energy DOF independently. A spin-analyzer in $+\hat z$ direction together with a $\pi/2$ spin-turner enabled the selection of neutrons in xy-plane. An accelerator coil, oriented in $B_{acc}$$+\hat z$ was used to adjust the spin phase $\alpha=0$, $\pi/2$, $\pi$, $3\pi/2$.

\begin{figure}
\begin{center}
   \includegraphics[width=0.45\textwidth]{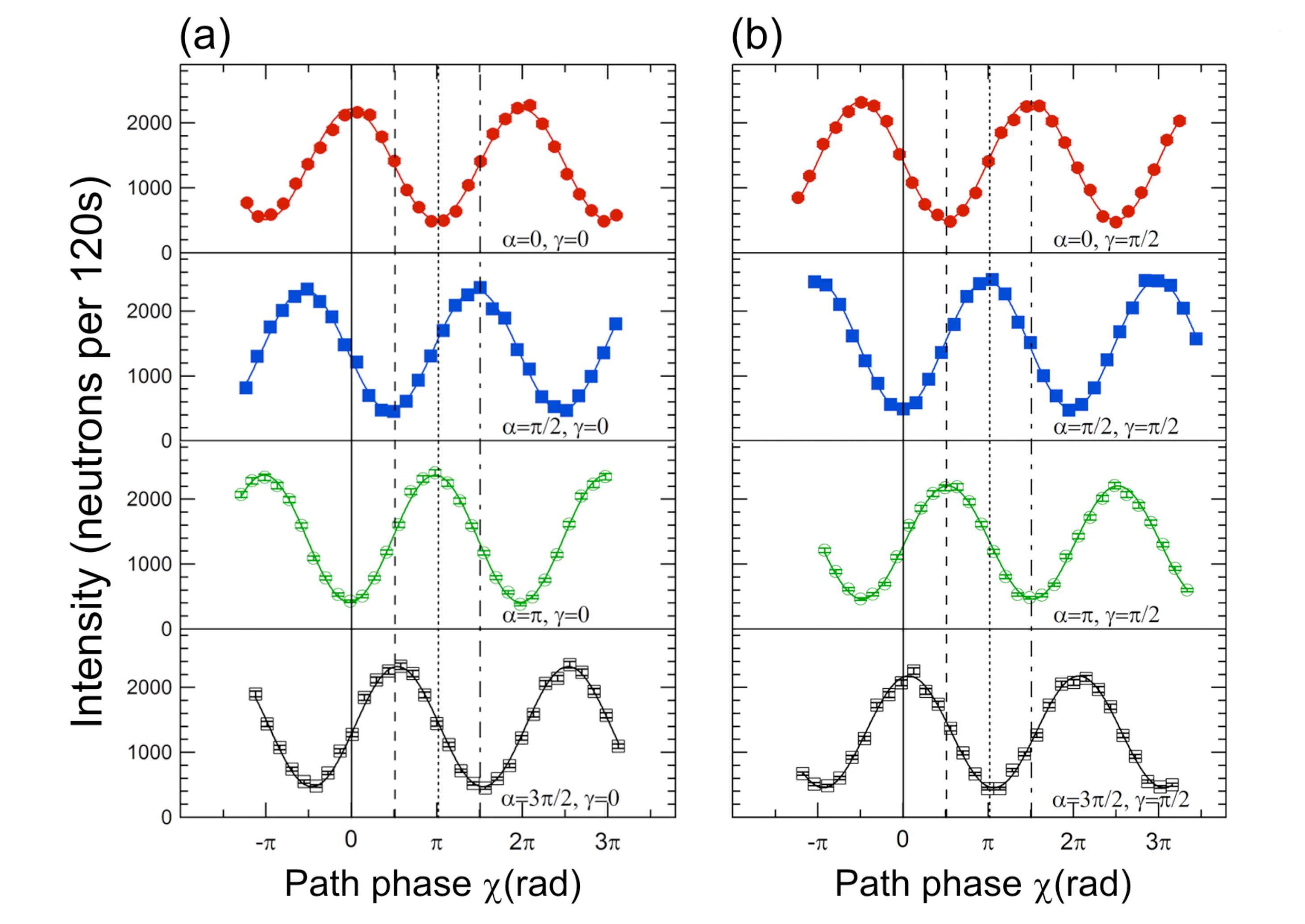}
   \caption{Typical intensity modulations obtained by varying the path phase $\chi$. The energy phase $\gamma$ is tuned at (a) 0 and (b) $\pi/2$. The spin phase $\alpha$ is set at 0, $\pi/2$, $\pi$,
   and $3\pi/2$ (from upper to lower panels).
   }
\label{Osci}       
\end{center}
\end{figure}

To determine the expectation values in $M$ in Eq. (\ref{eq:eq3}), we performed 16 independent path phase $\chi$ scans by tuning the spin phase $\alpha$, and the energy phase $\gamma$ each at $0$, $\pi/2$, $\pi$ and $3\pi/2$. Typical intensity modulations are depicted in Fig. \ref{Osci}. Obtained intensity oscillations were fitted to sinusoidal curves by the least squares method. The contrasts of the sinusoidal oscillations were just below $70\%$. From the intensities indicated by lines ($\chi=0$, $\pi/2$, $\pi$, $3\pi/2$), four expectation values were extracted. Statistical errors were estimated to $\pm0.001$ taking all fit-errors from single measurement curves into account. Four repeated measurements were summed up as weighted averages, then the final value and the error were determined. The final errors are the sum of systematic and statistical errors. We obtained four expectation values:
\begin{equation}
\left\{{\begin{array}{lll}
         E(\sigma^s_x\sigma^p_x\sigma^e_x)=+0.659(2) \\
         E(\sigma^s_x\sigma^p_y\sigma^e_y)=-0.632(2) \\
         E(\sigma^s_y\sigma^p_x\sigma^e_y)=-0.603(2) \\
         E(\sigma^s_y\sigma^p_y\sigma^e_x)=-0.664(2). \\
        \end{array}
} \right.
\end{equation}
From these values, the final $M$-value was calculated as
\begin{equation}
	M=2.558\pm0.004,
\end{equation}
which clearly exhibits a violation of the Mermin-like inequality, $|M|\leq2$, and confirms the invalidity of the assumption of non-contextuality. The deviation from the ideal value of 4 is solely due to the reduced contrast (just below $70\%$) of the interferograms.

\section{Concluding remarks}

Tripartite entanglement for single particles is investigated: in particular,
coherent and individual manipulations of the neutron's spin, path and energy DOFs are achived.
A GHZ-like state is implemented and used to confirm the fact that
a Mermin-like inequality is violated. This experiment clearly shows that
the total energy DOF is an appropriate two-level subsystem to be entangled with other DOFs.
It is worth noting here that, while the total-energy subsystem is treated here as a discrete two-level quantum system, the discreteness is not of natural origin but artificially created:
our experiment shows the possibility of introducing artificial discrete \emph{quantum} levels in a per se continuously distributed space.
This fact can provide flexible resources for quantum information and  communication applications; for instance, multi energy-levels in a single-particle system, created by sequential  energy manipulation schemes with multiple frequencies, can be exploited.
Furthermore, the concept of entanglement between different DOFs for single neutrons
can be extended, e.g., by using the k-vector (kinetic energy) as DOF which can be manipulated independently of the path, spin and the total energy by tuning an (external) magnetic field.
We are proceeding with experiments in this connection.



\section*{Acknowledgements}
We thank all colleagues who were involved in carrying out the experiments presented here.
This work has been supported partly by the Austrian Science Fund (FWF), No. P21193-N20 and Hertha-Firnberg-Programm T389-N16.



\bibliographystyle{elsarticle-harv}

\begin{thebibliography}{00}

\bibitem{EPR}
A. Einstein, B. Podolsky, and N. Rosen,
Phys. Rev. {\bf 47}, 777 (1935).

\bibitem{Bell}
J. S. Bell,
Physics {\bf 1}, 195 (1964).

\bibitem{WJSWZ98}
G. Weihs, T. Jennewein, C. Simon, H. Weinfurter, and A. Zeilinger,
Phys. Rev. Lett. {\bf 81}, 5039 (1998).

\bibitem{Matsukevich}
D. N. Matsukevich, P. Maunz, D. L. Moehring, S. Olmschenk, and C. Monroe,
Phys. Rev. Lett. {\bf 100}, 150404 (2008).

\bibitem{Mermin93}
N. D. Mermin,
Rev. Mod. Phys. \textbf{65}, 803 (1990).

\bibitem{Hasegawa03}
Y. Hasegawa, R. Loidl, G. Badurek, M. Baron, and H. Rauch,
Nature (London) \textbf{425}, 45 (2003).

\bibitem{Amselem09}
E. Amselem, M. Radmark, M. Bourennane, and A. Cabello,
Phys. Rev. Lett. \textbf{103}, 160405 (2009).

\bibitem{KS67}
S. Kochen and E. P. Specker,
J. Math. Mech. {\bf 17}, 59 (1967).

\bibitem{GHZ}
D.M. Greenberger, M.A. Horne, A. Zeilinger,
in M. Kafatos (Ed.), Bell's Theorem, Quantum Theory, and Conceptions of the Universe, Kluwer Academic, Dordrecht (1989), p. 69-72; see also arXiv:0712.0921.

\bibitem{GHSZ}
D. M. Greenberger, M. A. Horne, A. Shimony, and A. Zeilinger,
Am. J. Phys. \textbf{58}, 1131 (1990).

\bibitem{Mermin90b}
N. D. Mermin,
Phys. Rev. Lett. \textbf{65}, 1838 (1990).

\bibitem{Pan}
J. W. Pan, D. Bouwmeester, M. Daniell, H. Weinfurter, and A. Zeilinger,
Nature \textbf{403}, 515 (2000).

\bibitem{Zhao}
Z. Zhao, Y. A. Chen, A. N. Zhang, T. Yang, H. J. Briegel, and J. W. Pan,
Nature \textbf{430}, 54 (2004).

\bibitem{Sackett}
C. A. Sackett, D. Kielpinski, B. E. King, C. Langer, V. Meyer, C. J. Myatt, M. Rowe, W. A. Turchette, W. M. Itano, D. J. Wineland, and C. Monroe,
Nature \textbf{404}, 256 (2000).

\bibitem{Hasegawa2010}
Y. Hasegawa, R. Loidl, G. Badurek, K. Durstberger-Rennhofer, S. Sponar, and H. Rauch,
Phys. Rev. A \textbf{81}, 032121 (2010).

\bibitem{Mermin90a}
N. D. Mermin,
Am. J. Phys. \textbf{58}, 731 (1990).

\bibitem{Mermin90}
N. D. Mermin,
Phys. Rev. Lett. \textbf{65}, 3373 (1990).

\bibitem{Sponar08a}
S. Sponar, J. Klepp, R. Loidl, S. Filipp, G. Badurek, Y. Hasegawa, and H. Rauch,
Phys. Rev. A \textbf{78}, 061604(R) (2008).

\bibitem{Rauch00}
H. Rauch and S. A. Werner,
Neutron interferometry, Clarendon Press, Oxford (2000).

\bibitem{Summi93}
J. Summhammer, 
Phys. Rev. A \textbf{47}, 556 (1993).

\bibitem{Golub94}
R. Golub, R. G\"{a}hler, and T. Keller,
Am. J. Phys. \textbf{62}, 779 (1994).

\bibitem{Sponar08b}
S. Sponar, J. Klepp, R. Loidl, S. Filipp, G. Badurek, Y. Hasegawa, and H. Rauch,
Phys. Lett. A \textbf{372}, 3153 (2008).

\bibitem{Cabello2002}
A. Cabello,
Phys. Rev. A \textbf{65}, 032108 (2002).




 \end{thebibliography}



\end{document}